\documentclass[amssymb,prb,preprint,aps]{revtex4}
\usepackage{epsfig}

\begin{document}
\title{Superexchange in Dilute Magnetic Dielectrics: Application to
(Ti,Co)O$_2$}
\author{K.Kikoin}
\affiliation{Department of Physics, Ben-Gurion University, Beer
Sheva 84105, Israel}
\author{V. Fleurov\footnote{Email:fleurov@post.tau.ac.il.}}
\affiliation{Raymond and Beverly Sackler Faculty of Exact Sciences,
School of Physics and Astronomy, Tel-Aviv University, Tel-Aviv 69978
Israel}
 \affiliation{Laue-Langevin Institute, F-38042, Grenoble, France}
\begin{abstract}
We extend the model of ferromagnetic superexchange in dilute
magnetic semiconductors to the ferromagnetically ordered highly
insulating compounds (dilute magnetic dielectrics). The intrinsic
ferromagnetism without free carriers is observed in oxygen-deficient
films of anatase TiO$_2$ doped with transition metal impurities in
cation sublattice. We suppose that ferromagnetic order arises due to
superexchange between complexes [oxygen vacancies + magnetic
impurities], which are stabilized by charge transfer from vacancies
to impurities. The Hund rule controls the superexchange via empty
vacancy related levels so that it becomes possible only for the
parallel orientation of impurity magnetic moments. The percolation
threshold for magnetic ordering is determined by the radius of
vacancy levels, but the exchange mechanism does not require free
carriers. The crucial role of the non-stoichiometry in formation of
the ferromagnetism makes the Curie temperatures extremely sensitive
to the methods of sample preparation.
\end{abstract}

\maketitle

\section{Introductory notes}

Recent experimental and theoretical investigations of dilute
ferromagnetic semiconductors are concentrated mostly on II-VI and
III-V materials doped with Mn impurities (see, e.g., \cite{Avsh}).
The theoretical explanation is based on the Vonsovskii-Zener model,
which implies existence of a direct exchange interaction between the
localized spins of magnetic ions (Mn) and itinerant spins of free
carriers (holes in case of (Ga,Mn)As and related
materials).\cite{Dietl} This interaction generates RKKY-type
indirect exchange between Mn ions, and the latter is claimed to be
the source of long-range FM order in dilute magnetic semiconductor
(DMS) alloys. It was shown recently that the kinematic exchange
interaction (specific version of superexchange between localized Mn
moments via empty valence states of the host material) arises in
$p$-type III-V DMS.\cite{Krst}  This mechanism  works together with
the RKKY interaction because of the noticeable hybridization between
d-electrons of Mn ions and p-holes near the top of the valence band.
Both mechanisms are characterized by a direct proportionality
between the carrier concentration and Curie temperature $T_C$.
However, it was pointed out recently\cite{BBZ06} that an application
of the RKKY based model to disordered DMS may be questionable if one
properly takes oscillating character of the RKKY interaction into
account.

Meanwhile, another family of dilute ferromagnetic alloys was
discovered during the recent five years, where the ferromagnetic
order with a high $T_C$ exists in spite of the low carrier
concentration. The ferromagnetic order was observed even in the
insulating materials with carriers frozen out at $T\to 0$. These are
metal oxides doped with transition metal ions (see, e.g., Ref.
\onlinecite{Peart} for a review). The absence of free carriers
leaves no room for a RKKY interaction in this case, so the question
about the origin of ferromagnetism arises anew.

The metal oxides like ZnO, SnO$_2$, TiO$_2$ are classified as
wide-gap semiconductors. In this sense they are close to the
wide-gap III-V nitrides GaN and AlN. We analyzed the case of
$n$-(Ga,Mn)N in Ref. \onlinecite{Krst}, where the leading
interaction mechanism is the Zener type double exchange \cite{Zener}
vie empty states in the impurity band. Magnetic ions (MI) in all
these compounds substitute for metallic cations. From this point of
view, the difference between the two groups is in the charge state
of substitutional impurity: the neutral state of an impurity should
be MI$^{3+}$ in nitrides and MI$^{4+}$ in oxides. This difference
implies different structures of spin multiplets, which is important
for the magnetic properties of metallic alloys. However, due to a
noticeable covalency of the host materials, MI should saturate the
broken bonds by its own valent electrons, which are donated by the
3d shells. This means that the violation of the electronic structure
inserted by the dopants cannot be ignored in the studies of magnetic
properties of DMS. Moreover the recent paper \cite{klg06} emphasizes
the role of bound exciton states in ZnO bases compounds and
indicates the correlation between between the experimentally
observed chemical trends in ferromagnetism and the electron binding
energies to various MI.

Among the most salient features of magnetism in dilute ferromagnetic
oxides one should mention an extreme sensitivity of the magnetic
order to the growth and annealing conditions.\cite{Peart,Coey} We
believe that this is an integral feature of magnetism in these
materials, and the ferromagnetic ordering with high $T_C$ is
mediated by intrinsic or extrinsic defects, which form complexes
with magnetic dopants. Such point of view is supported by recent
experimental studies of Co- doped\cite{Grif} and Cr-doped\cite{Cr}
TiO$_2$. It was noticed, in particular, that in the most perfect
${\rm (Ti,Cr)O_2}$ crystals the hysteresis loop at a given
temperature is essentially less distinct than in 'bad quality'
samples.\cite{Cr} Besides, the enormous scatter of effective
magnetic moment per cation is observed in all ferromagnetic oxides,
and the concentration of magnetic dopants lies far below the
percolation threshold $x_c$ associated with the nearest-neighbor
cation coupling.\cite{Coey} Basing on these facts, one concludes
that non-magnetic defects are also involved in formation of the
localized magnetic moments. They influence their magnitude and
localization, and this influence is sensitive to the growth and
annealing regimes.

Recent theories of magnetism in $n$-type dilute magnetic oxides
appeal to magnetic polarons as mediators of indirect exchange
between magnetic dopants.\cite{Coey,polar} In this case the {\em
extrinsic} defects are donor impurities, which donate free electrons
occupying the bottom of conduction band. These electrons form
shallow spin-polarized polaronic states due to strong exchange with
the localized magnetic moments of transition metal dopants. A large
radius of these states makes the polaronic percolation threshold
essentially lower than $x_c$.

One should note, however, that the carrier concentration in these
compounds is vanishingly small, and that is why they have been
proposed to be called dilute magnetic dielectrics (DMD).\cite{Grif}
It means that we need now to find an exchange mechanism, which works
in the case of really insulating oxides. In this paper we construct
a model of insulating oxide, which considers {\em intrinsic}
non-magnetic defects as an integral part of the ordering mechanism.
We concentrate mainly on the superexchange between magnetic ions
mediated by oxygen vacancies, which is believed to be the source of
ferromagnetic order in TiO$_2$ diluted with Co.\cite{Grif}

\section{Model of dilute magnetic oxide}

As has been established three decades ago, \cite{Haldane,Fleurov1}
the basic microscopic Hamiltonian, which gives an adequate
quantitative and qualitative description of electronic and magnetic
properties of semiconductors doped by transition metal ions is the
Anderson Hamiltonian \cite{And} (see summarizing monographs
\cite{Kikoin,Zunger}). Indirect RKKY \cite{Mcdon}, superexchange and
double exchange \cite{Krst} interactions between magnetic ions may
be derived microscopically from the multisite generalization of the
Anderson model.\cite{Alex} The Hamiltonian of this model reads
\begin{equation}
{\sf H} ={\sf H}_h +{\sf H}_d +{\sf H}_{hd},  \label{And1}
\end{equation}
where ${\sf H}_h$ describes the electronic structure of the host
semiconductor, ${\sf H}_d = \sum_j{\sf H}_{dj}$ is the Hamiltonian
of strongly interacting electrons in the $d$-shells of magnetic ions
located in the lattice sites $j$, and ${\sf H}_{hd} = \sum_j {\sf
H}_{hdj}$ is the hybridization Hamiltonian describing the covalent
bonding between the impurity $3d$-electrons and the host $p$
electrons (the contribution of hybridization with the host $s$
electrons in the effective exchange is generally negligibly small).
Usually the direct $dp$ exchange is neglected in the Anderson model,
since the indirect exchange coupling dominates in the covalent
materials.\cite{Zunger,Kikoin}

From the point of view of the Anderson model, the crucial difference
between the narrow-gap and wide-gap semiconductors is in the
position of d-levels of the impurity electrons relative to the top
of the valence band. Due to a strong electron-electron interaction
in the $3d$ shell (Coulomb blockade), the impurity energy level
$\varepsilon_d$ should be defined as an "addition energy" for the
reaction $d^{n-1}+ e_b \to d^{n}$, i.e. the energy cost of adding
one more electron, $e_b$, from the band continuum to the impurity
3d-shell,
\begin{equation}\label{addit}
\varepsilon_d = E(d^{n}) - E(d^{n-1})
\end{equation}
where $E(d^{p})$ is the total energy of a "pseudo-ion" consisting
of an impurity in a configuration $3d^p$ and distorted electron
distribution in the host semiconductor \cite{Kikoin}.

In this definition the energy is counted from the top of the
valence band, which is the source of additional electrons for the
impurity. It is known, e.g. that Mn, the most popular magnetic
impurity, has especially deep $d$ level corresponding to the
half-filled $d$ shell in the configuration $3d^5$. This level is
always occupied in such narrow-band semiconductors as (Ga,Mn)As,
(Ga,Mn)P, (In,Mn)As. This means that the addition energy
$\varepsilon_d = E(d^5) - E(d^4)$ is well below the energy,
$\varepsilon_{p}$, of the top of the valence band. On the
contrary, $\varepsilon_d > \varepsilon_{p}$ in the wide gap
semiconductors like GaN,\cite{Zung04} and the occupation of this
state depends on the type and concentration of other defects.

One should compare this picture with the energy spectra of
magnetically doped metal oxides. Situation with the transition metal
impurities in ZnO is apparently close to that in GaN. Both these
materials are wide band gap semiconductors, and the only difference
is that the neutral state of magnetic impurity in ZnO is MI$^{2+}$.
The corresponding energy levels arise, as a rule, within the band
gap, and the states with other oxidation numbers may appear only due
to an additional $n$ - or $p$ - doping.

The calculation\cite{Mizu} of addition energies for MI$^{n+}$
states in rutile TiO$_2$ has shown that the "energy levels"
$\varepsilon_d$ for neutral MI$^{4+}$ states are usually deep in
the valence band for heavy transition metal ions (starting with
Mn), whereas the condition $\varepsilon_d > \varepsilon_{p}$ is
fulfilled for the addition energies of charged MI$^{3+}$ and
MI$^{2+}$ ions. Recent LSDA+U calculations \cite{Park} for anatase
TiO$_2$ doped with Mn, Fe and Ni are in a general agreement with
those earlier cluster calculations.

This means that the ground state of this system should be
TiO$_2$:MI$^{4+}$. In this situation the Anderson Hamiltonian may be
immediately transformed into the $sd$-exchange model by means of the
Schrieffer-Wolff transformation. In the absence of free carriers the
short-range effective inter-impurity exchange interaction is
exponentially weak and there is no chance for magnetic ordering with
reasonably high $T_C$. However, the situation in real materials is
quite different. Leaving aside early results for the samples with
secondary phase inclusions,\cite{Peart} we concentrate on the recent
data for the materials, which are believed to be genuine
ferromagnetic DMD free from magnetic precipitates. As was mentioned
above, the room temperature ferromagnetic order was detected in Cr
and Co doped anatase TiO$_2$, and in both cases native defects are
involved in formation of the ferromagnetic order.\cite{Cr,Grif} In
the latter case these defects are apparently oxygen vacancies.
Below, we adapt our theory\cite{Krst} for vacancy mediated indirect
exchange. The case of (Ti,Cr)O$_2$ will be discussed in the
concluding section.

\section{Vacancy-related superexchange mechanism}

In the highly insulating ferromagnetic (Ti,Co)O$_2$,  magnetic ions
are in Co$^{2+}$ state.\cite{Grif} A seeming contradiction with the
requirement of electrical neutrality of the substitutional
impurities\cite{Mizu} may be resolved, if one takes into account the
intrinsic non-stoichiometry of the samples: oxygen vacancies V$_O$,
are created in the process of sample preparation. It follows from
the general neutrality consideration that an oxygen vacancy binds
two electrons on a discrete level in the band gap to saturate the
dangling bonds. In the process of annealing a noticeable amount of
oxygen vacancies are captured by Co substitution impurities in
cation sites. In the case, when ${\rm V_O}$ arises in the octahedron
containing a Co ion, the double defect ${\rm Co-V_O}$ is formed, and
its stability (or metastability with respect to formation of ${\rm
Ti-V_O}$ complexes) is determined by the charge transfer of these
two electrons in accordance with the reaction
\begin{equation}\label{vac}
{\rm V_O^0}(p^2) + {\rm Co}^{4+}(d^5) \to [{\rm V_O^{2+}}{\rm
Co}^{2+}(d^7)].
\end{equation}
A presumable scheme of the energy levels, which substantiates such
charge transfer is shown in Fig. 1. The configurations $d^n$ of
$3d$ shells corresponding to different charge states of Co and Cr
impurities are pointed out. The dashed band shows the position of
vacancy-related levels below tho bottom of conduction band
$\varepsilon_c$.
\begin{figure}[htb]
\centering
\includegraphics[width=160mm,angle=0]{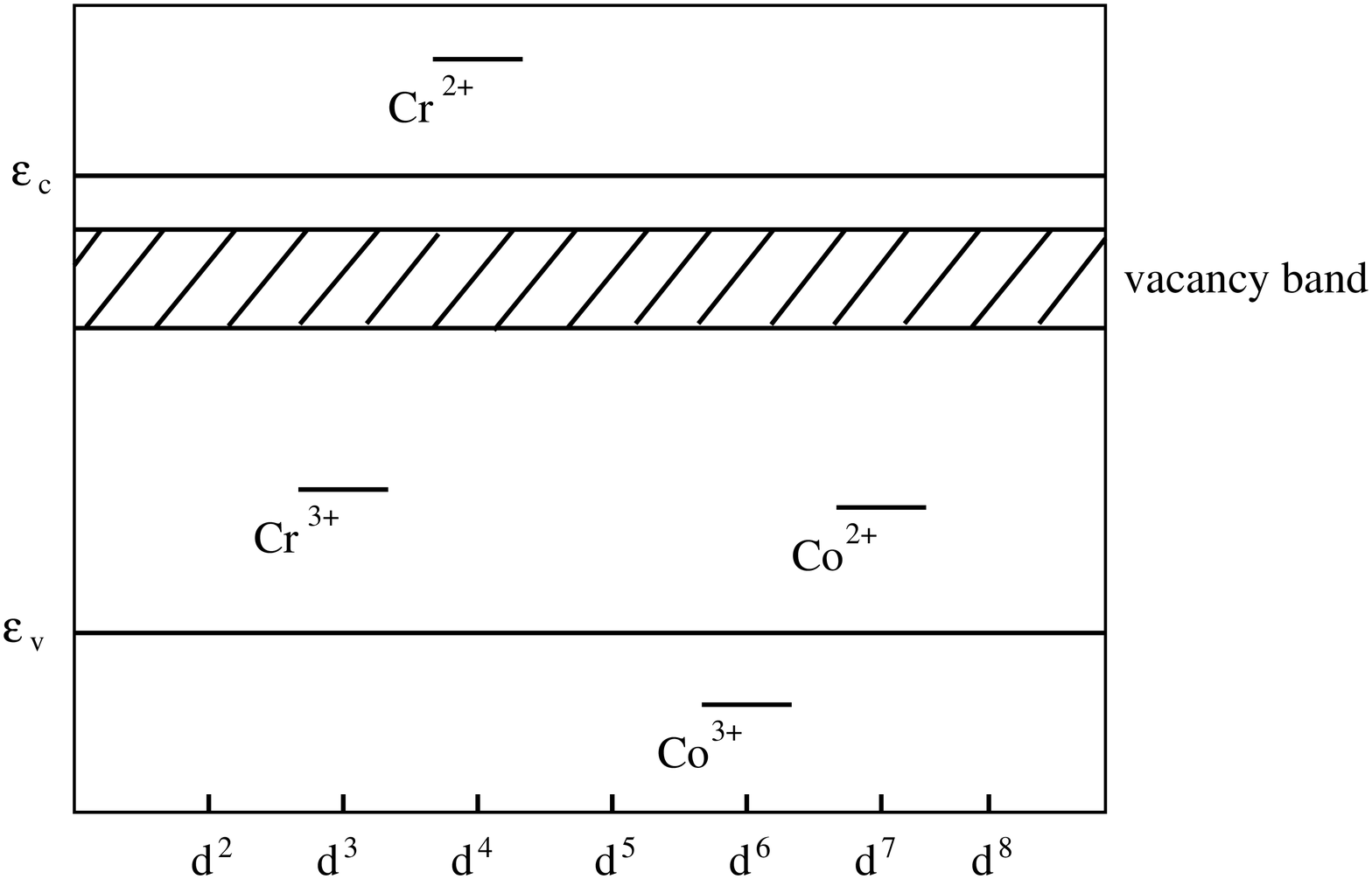}
\caption{Impurity-related and vacancy-related energy levels in
cation substituted non-stoichiometric TiO$_2$.} \label{lev}
\end{figure}

As a result of charge transfer reaction (\ref{vac}), the vacancy
related electrons are captured by $[{\rm V_O^{2+}}{\rm Co}^{2+}]$
complexes, the system remains insulating and the magnetic
interaction between these complexes predetermines the magnetic
properties of the system.

In this section we construct a microscopic model of impurity-vacancy
complexes basing on our previous studies\cite{Krst} of double
ferromagnetic exchange in (Ga,Mn)As and the electronic structure of
complexes Cr-V$_O$ in (Ga,Cr)As,\cite{Perv} as well as on the
numerical calculations of the electronic structure of ${\rm V_O}$
defects in ${\rm TiO_2}$~\cite{Halley} and doped non-stoichiometric
(oxygen deficient) $\rm (Ti,Co)O_2$~ \cite{Park,Anisim}.

The structure of Hamiltonian, which describes the oxygen deficient
(Ti,Co)O$_2$ is the same as in Eq. (\ref{And1}), but the oxygen
vacancies should be taken into account in the host Hamiltonian
$H_h$. We start with a quantum-chemical picture of an isolated
defect cell, which contains both Co impurity substituting for a Ti
cation and the vacancy in one of the apices of oxygen octahedra. We
mark the cell by the index $j$ and consider the Hamiltonian
\begin{equation}
\label{mod} {\sf H}_{Vj} ={\sf H}'_{hj} +{\sf H}_{dj} +{\sf
H}_{hdj}.
\end{equation}
Now the first term ${\sf H}'_{hj}$ describes the host crystal with
the vacancy, ${\rm V_O}$, in the cell $j$. This Hamiltonian may be
diagonalized, and the resulting spectrum contains the band continuum
distorted by the vacancy potential and the discrete levels
$\varepsilon_{v\mu}$ for electrons bound on the dangling bonds in
the defect cell.

The Hamiltonian ${\sf H}_{hdj}$ rewritten in this basis contains
terms describing hybridization with both continuum and discrete
states of ${\sf H}'_{hj}$. The general form of ${\sf H}_{hdj}$ is
\begin{eqnarray}\label{hyb}
{\sf H}_{hdj} & = & \sum_{\mu\nu\sigma} \sum_{\tau_d\tau_o}
\left(V_{j\nu\mu} (\mbox{\boldmath $\tau$}_d,
\mbox{\boldmath$\tau$}_o)d^\dag_{j + \tau_d, \mu\sigma} c_{j +
\tau_o,\nu\sigma} + h.c.\right)\nonumber
\\
&+& \sum_{\kappa \mu\sigma}\sum_{a=v,c} \left(V_{ja\mu}(\mbox{
\boldmath$\tau$}_d, \kappa) d^\dag_{j + \tau_d,\mu\sigma}c_{\kappa
a\sigma} + h.c.\right).
\end{eqnarray}
Here the operators $d_{j+\tau_d,\mu\sigma}$ and $c_{j + \tau_o,
\nu\sigma}$ stand for the localized $d\mu$ - and $p\nu$ - orbitals
on the impurity site ${\bf R}_i= {\bf R}_j + \mbox{ \boldmath$
\tau$}_d$ and the vacancy cite ${\bf R}_o = {\bf R}_j + \mbox{
\boldmath$ \tau$}_o$, respectively, and the operators $c_{\kappa
a\sigma}$ describe the continuous states in the host crystal valence
$(a = v)$ and conduction $(a = c)$ bands.

Information about the vacancy-related states in anatase TiO$_2$ is
rather scanty,\cite{Weng} so we refer to the data available for
oxygen-deficient rutile TiO$_2$. In accordance with numerical
calculations\cite{Halley,Anisim}, which correlate with the
experimental observations, V$_O$ creates donor levels
$\varepsilon_o$ a few tenth of an eV below the conduction band. The
donor electrons saturate the dangling bonds with the neighboring
atoms, so that the defect wave function has the largest amplitude at
the next Ti neighbor of the vacancy and extends to several
coordination spheres. The extended defect states form a band of
donor levels already for 1\% concentration of oxygen vacancies, i.e.
the wave functions of the vacancy states effectively overlap when
the vacancies are separated by a distance of about 5 interatomic
spacings. This observation will be used below when estimating the
Curie temperature. The calculated electronic structure for Co-doped
oxygen deficient anatase TiO$_2$ gives similar picture for the
charge distribution and the density of states.\cite{Weng}

This observation verifies the model Hamiltonian (\ref{hyb}) for the
Co substitutional impurities. Its form implies a strong
nearest-neighbor hybridization between the vacancy and impurity
related states. Since the conduction band of TiO$_2$ is formed
mainly by the $d$-states of Ti sublattice, the matrix elements
$V_{c\mu}$ in the second term in this Hamiltonian describe the
hybridization of the Co impurity $d$-states with those in the
cationic sublattice. Apparently one may neglect the hybridization
$V_{c\mu}$ in comparison with $V_{j\nu\mu}$, and retain in this term
only the hybridization $V_{jv\mu}$ with the oxygen-related p-states
in the valence band.

We return now to are the energy level scheme Fig. \ref{lev}
substantiating the charge transfer reaction (\ref{vac}). It is
emphasized that we compare the one-electron energy levels of the
oxygen vacancy with the corresponding levels of Co impurities in
different charge states. These levels are defined as addition
energies (\ref{addit}) corresponding to recharging processes ${\rm
Co}^{m+}/{\rm Co}^{(m-1)+}$. These energies may be taken, i.g. from
Ref. \onlinecite{Mizu}. Hybridization $V_{j\nu\mu}$ allows for an
electron transfer of the two electrons from the vacancy level first
to the level $\varepsilon_d(3+/4+)$ and then to the level
$\varepsilon_d(2+/3+)$ and formation of the complex defect $[{\rm
V_O^{2+}}{\rm Co}^{2+}]_j$ with seven electrons occupying the
3d-shell of the ion ${\rm Co}^{2+}$ and empty extended V$_{\rm
O}$-related states.

Indirect exchange interaction arises when the wave functions of two
complexes centered in the cells $j$ and $i$ overlap. It is clear
that the overlap is controlled by the radius of extended V$_{\rm
O}$-related state. An obvious interaction mechanism is the
superexchange, where two magnetic impurities exchange electrons via
empty vacancy levels. It will be shown below that this interaction
favors ferromagnetic ordering in the system.

\begin{figure}[htb]
\centering
\includegraphics[width=60mm,angle=0]{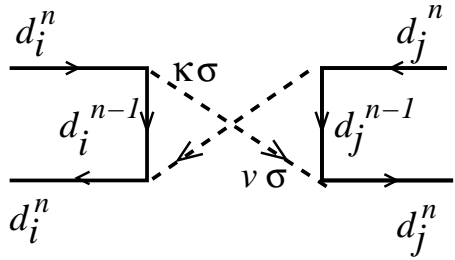}
\caption{Diagrammatic representation of the superexchange
interaction between two magnetic impurities in charge states
$d_i^n$ and $d_j^n$ via the intermediate states $d_i^{n-1}$ and
$d_j^{n-1}$ and empty levels $p\sigma$ (see text for further
explanation).} \label{zen}
\end{figure}

The simplest way to construct an effective superexchange interaction
operator is to project the two-impurity Hamiltonian ${\sf H}_{ji} =
{\sf H}_{dj} + {\sf H}_{di} +{\sf H}^\prime_{hji} + {\sf H}_{hdj}
+{\sf H}_{hdi}$ onto the subspace $\langle ji|\ldots
|ji\rangle_{2+}$ with fixed ${\rm Co^{2+}}$ configurations of the
two magnetic ions and also project out the charged impurity states
${\rm Co^{3+}}$, which arise due to the hybridization ${\sf H}_{hdj}
+ {\sf H}_{hdi}$. Here ${\sf H}^\prime_{hji}$ is the host
Hamiltonian with two oxygen vacancies in the cells $j,i$. Then the
fourth order term in the impurity-vacancy hybridization
\begin{equation}\label{vert}
\langle ji|{\sf H}_{hdj}\frac{1}{E_0 - {\sf H}_{ji}^{0}}{\sf
H}_{hdi} \frac{1}{E_0 - {\sf H}_{ji}^{0}} {\sf H}_{hdi}\frac{1}{E_0
- {\sf H}_{ji}^{0}} {\sf H}_{hdj}|ji\rangle_{2+}
\end{equation}
gives the effective exchange vertex, represented graphically in Fig.
\ref{zen}.
The resolvents, which appear in (\ref{vert}) correspond to the
ground state $E_0$ of the non-hybridized Hamiltonian ${\sf
H}^{0}_{ji} = {\sf H}_{dj} + {\sf H}_{di} + {\sf H}^\prime_{hji}$.
The solid lines in this diagram stand for the definite charge and
spin state of the impurity in the site $i$ or $j$ (Co$^{2+}(d^7)$
and (Co$^{3+}(d^6)$ in a given spin and orbital configuration).
The vertices correspond to various terms in the hybridization
Hamiltonian (\ref{hyb}), which changes the impurity state from
$d^7_i$ to $d^6_i$ and v.v. In the vertex $V_{v\mu}$ an electron
appears on the {\em empty} vacancy $\nu$ - orbital with the spin
$\sigma$. In the vertex $V_{a\mu}$ a hole (electron) appears in
the valence (conduction) band. Yet, a similar process $d_j^7 \to
d_j^6 + \nu'\sigma$ occurs around the counterpart site $j$. Then
the two p-electrons change their 'hosts' (two dashed lines in Fig.
\ref{zen}), and this is the end of the superexchange process. This
scheme allows one to roughly estimate the effective exchange
coupling as
\begin{equation}\label{estim}
J_Z \sim V^4/\Delta_{do}^2 D_o,
\end{equation}
where $\Delta_{do}$ is the energy of the  transition (\ref{vac}),
and $D_o$ is the characteristic energy scale of the vacancy related
band under the bottom of conduction band.

A similar estimate may be found, e.g. in Refs. \onlinecite{Khom},
where the superexchange between singly occupied states in the
$4f$-shell via empty states in conduction band was studied ($D_o$ in
that case is the conduction band width). In the case of a singly
occupied shell $f^1$ or $d^1$ there is no Hund interaction in the
unfilled shell. As a result antiferromagnetic part of this
interaction prevails (ferromagnetic channels are blocked by the
Pauli principle).\cite{Khom} In our case the superexchange mechanism
favors the ferromagnetic alignment of spins since the intrashell
Hund interaction suppresses the exchange by the electrons with
antiparallel spins (see next section).

There seems to be much in common between the vacancy-related
mechanism represented by Eqs. (\ref{vert}), (\ref{estim}) and the
spin-polaron mechanism offered in Ref. \onlinecite{Coey}. In both
cases the effective radius of magnetic interaction is controlled by
the extended donor states. It should be emphasized, however, that
unlike the indirect exchange via the states in the donor impurity
band\cite{Coey}, in our case the band of vacancy states may be empty
(provided all the electrons from the vacancy levels are captured by
acceptor-like magnetic impurities), so that this mechanism may be
realized in highly insulating materials. On the other hand, the
empty donor band of extended vacancy related states plays, in oxygen
deficient (Ti)O$_2$, the role similar to that played by the empty
states in the hole pockets of $p$-type (Ga,Mn)As in superexchange
between magnetic ions (see Ref. \onlinecite{Krst}). We will return
to the comparison of superexchange mechanism in various dilute
magnetic compounds in the concluding section.

\section{Microscopic theory of superexchange in DMD}

The simple estimate (\ref{estim}) by means of the fourth order
perturbation equation (\ref{vert}) may be essentially improved.
Basing on the theory of {\em isolated} transition metal impurities
in semiconductors,\cite{Kikoin,Zunger} one may consider the effects
of single-site hybridization between the $d$-orbitals of impurity
and oxygen-related states in the valence band and vacancy levels and
then use the perturbation theory approach for the {\em intersite}
hybridization effects responsible for the superexchange.

It is  known that the main result of this hybridization is
'swelling' of the wave functions of the impurity
$d$-electrons.\cite{Kikoin,Zunger} These functions acquire 'tails'
formed by a superposition of Bloch waves from valence and conduction
bands. A localized vacancy state is also a superposition of the same
Bloch waves. All these covalent effects modify the simple estimate
(\ref{estim}), which necessitate a more refined analysis within the
microscopic Anderson-like Hamiltonian.

In this section we will derive equations for the superexchange
interaction in the system of defects $[{\rm Co_{Ti}, V_O}]$  by
means of the Green function method developed in Ref.
\onlinecite{Krst}. For the purpose of practical calculations within
this method, it is more convenient to write the term ${\sf H}'_{hj}$
in the model Hamiltonian (\ref{mod}) in the form
\begin{equation}\label{band}
{\sf H}'_{hj} = \sum_{\kappa a\sigma} \varepsilon_\kappa
c^\dag_{\kappa a\sigma} c_{\kappa a\sigma} + \sum_{\nu\sigma}
\varepsilon_{v\nu} c^\dag_{j + \tau_v,\nu\sigma}
c_{j+\tau_v,\nu\sigma}.
\end{equation}
It is assumed in Eq. (\ref{band}) that the host Hamiltonian with
vacancies has been diagonalized, and the band states are classified
by the index $\kappa$. Each vacancy in the site $j+\tau_v$ creates a
localized level in the gap. It was found in the previous numerical
calculations \cite{Mizu,Halley} that both the oxygen-related
$p$-states in the valence band and the titanium-related $d$-states
from the valence band contribute to formation of localized vacancy
states, and the vacancy-related level arises slightly below the
bottom of conduction band.

The term ${\sf H}_d$ describing the subsystem of magnetic
impurities in the Hamiltonian (\ref{And1}) has the form
\begin{equation}\label{dot}
{\sf H}_d = \sum_j{\sf H}_{dj} = \sum_j \sum_{\mu\sigma}
\varepsilon_{\mu} d^\dag_{j\mu\sigma} d_{j\mu\sigma} + {\sf
H}_{j}^{corr}.
\end{equation}
Here the site indices are introduced in accordance with the
definition given in (\ref{hyb}), $H_{i}^{corr}$ includes all the
Coulomb and exchange interactions responsible for the Hund rule
within the $3d$-shells of Co ions. In accordance with this rule the
Co d-shell contains seven electrons, five of which form the closed
'inert' $d^5_\uparrow(t^3_{2}e^2)$ subshell and the remaining two
electrons form the open 'active' $ d^2_\downarrow (t_{\mu}e_{\mu'})$
subshell. Only the electrons in the open subshell are involved in
the hybridization induced superexchange. The reason for this
discrimination is the above mentioned swelling of the impurity
electron wave functions due to the hybridization with the host
electrons, which is stronger for the open subshell than for the
closed one.\cite{Kikoin,Zunger}

It is easily seen from (\ref{vert}) that the fourth order transition
$$
\langle d^5_{j\uparrow} d^2_{j\downarrow}, d^5_{i\uparrow}
d^2_{i\downarrow} |d^\dag_{i\downarrow}c_{\nu'\downarrow}{\sf
R}_{ji} d^\dag_{j\downarrow} c_{\nu\downarrow} {\sf R}_{ji}
c^\dag_{\nu\downarrow} d_{i\downarrow} {\sf R}_{ji}
c^\dag_{\nu'\downarrow} d_{j\downarrow} | d^5_{j\uparrow}
d^2_{j\downarrow},d^5_{i\uparrow} d^2_{i\downarrow}\rangle
$$
for ferromagnetically aligned spins of two impurities is possible,
whereas the similar process
$$
\langle d^5_{j\uparrow} d^2_{j\downarrow},d^5_{i\downarrow}
d^2_{i\uparrow}| d^\dag_{i\downarrow} c_{\nu'\downarrow}{\sf R}_{ji}
d^\dag_{j\uparrow}c_{\nu\uparrow} {\sf R}_{ji}
c^\dag_{\nu\uparrow}d_{i\uparrow}{\sf R}_{ji}
c^\dag_{\nu'\downarrow} d_{j\downarrow} |d^5_{j\uparrow}
d^2_{j\downarrow}, d^5_{i\downarrow} d^2_{i\uparrow}\rangle
$$
is suppressed due to the Hund rule for the occupation of subshells.
Here ${\sf R}_{ji} = (E_0 - {\sf H}_{ji}^{0})^{-1}$ is the zeroth
order resolvent.

Our task is to calculate the energy gain due to the electron
exchange between the impurities with parallel spin alignment.
Since no spin-flip occurs in the course of the exchange, the spin
index may be omitted in the Hamiltonian (\ref{hyb}), (\ref{band}),
(\ref{dot}). This energy gain is given by the following equation
\cite{Krst,Hald}
\begin{equation}\label{energy}
E^{magn} = \frac{1}{\pi}{\rm Im}\int^\infty_{-\infty}\varepsilon
{\rm Tr} \Delta {\sf G}[\varepsilon - i\delta\ {\rm
sign}(\varepsilon - \eta)] d\varepsilon,
\end{equation}
where $\Delta {\sf G}$ is the part of the single electron Green
function related to the spin-dependent interaction of magnetic
impurities,
$$
\Delta {\sf G}(z)= (z - {\sf H})^{-1} - (z - {\sf H}'_{hj})^{-1},
$$
$\eta$ is the chemical potential. The integral in (\ref{energy})
contains contributions from both the valence band continuum and
occupied discrete states in the energy gap.

When calculating $ \Delta {\sf G}(z)$, we exploited the following
features of our problem. \noindent
\\ (i) Since we are interested in
the energy gain due to the parallel spin ordering, the spin-flip
processes do not contribute to the relevant part of the system of
Dyson equations for the Green function and these equations allow for
an exact solution.\cite{Kikoin} \noindent
\\
(ii) The calculation procedure may be radically simplified by using
the analytical properties of ${\rm Tr} \Delta {\sf G}(z)$, which
allow one to calculate the energy by summing over the empty states
instead of integrating over the whole occupied part of the energy
spectrum.\cite{Krst} \noindent
\\
(iii) We are interested only in the trace of the Green function
defined in the subspace $\langle d_j^7 d_i^7|\ldots |d_i^7 d_j^7
\rangle$ (see Eq. (\ref{vert})).

As was discussed above, the leading contribution to the formation of
complex defect is due to the hybridization between the d-shells of
magnetic impurities and the empty states in the band of vacancy
related states below the bottom of conduction band. Thus, our next
task is to calculate the upward shift of the $V_{\rm O}$-related
levels. For this sake one has to find the poles of the single
electron Green function ${\sf G}$. This function has the following
block structure
\begin{equation}\label{matr1}
{\sf G} = \left(
\begin{array}{ccc}
{\sf G}_{\mu\mu'} &  {\sf G}_{\mu v'} & {\sf G}_{\mu\nu'} \\
{\sf G}_{v\mu'} & {\sf G}_{vv'} & 0 \\
{\sf G}_{\nu\mu'} & 0 & {\sf G}_{\nu\nu'}
\end{array}
\right)
\end{equation}
where the diagonal blocks are the two impurity (${\sf
G}_{\mu\mu'}$), band (${\sf G}_{vv'}$) and single-vacancy (${\sf
G}_{\nu\nu'}$) Green functions, respectively, and the off-diagonal
blocks stand for two types of hybridization. Only valence band
states are taken into account in the band component.

As shown above the exchange energy $E^{magn}$ is given  by the
shifts of the V$_O$ levels which are determined by the poles of {\sf
G}. It is sufficient to study the block ${\sf G}_{\nu\nu}$ only
since it has the same poles. Moreover, as follows from Eq.
(\ref{energy}), we need only its diagonal part $\widetilde
G_{\nu\nu}$ in order to calculate its trace. It reads
\begin{equation}\label{Dyson1}
\widetilde G_{\nu\nu} = g_\nu + g^2_\nu \sum_{i=1,2;\ \mu}
V^*_{\nu\mu}({\bf R}_i) G_{i\mu,i\mu} V_{\nu,\mu}({\bf R}_i) +
$$$$
g^2_v(\varepsilon) \sum_{i,j=1,2;\ \mu\mu',\ v'} V^*_{\nu\mu}({ \bf
R}_i) G_{i\mu,i\mu}(\varepsilon) V_{\nu',\mu}({\bf R}_i) g_{v'}(
\varepsilon) V^*_{\nu'\mu'}({\bf R}_j) G_{j\mu',j\mu'}(\varepsilon)
V_{ \nu,\mu'}({\bf R}_j).
\end{equation}
We have kept the leading terms in the hybridization of the vacancy
and impurity states contributing to the intersite exchange.
Hybridization of the empty conduction band states and the impurity
states is neglected in calculation of this exchange (see discussion
after Eq. (\ref{hyb}) in Section III). Since the spin-flip processes
are not involved in the calculation of $E^{magn}$, the equations may
be solved for each spin projection separately, and we omit below the
spin index for the sake of brevity.

We represent the impurity Green function in terms of the
irreducible representation of the symmetry group of axial defect
$[{\rm Co_{Ti}, V_O}]$, so that the impurity Green functions are
diagonal in this representation

\begin{equation}\label{Dyson2}
G_{d;i\mu,i\mu} = [\varepsilon - \varepsilon_{d;i\mu} -
\Sigma_{d;i\mu}(\varepsilon)]^{-1}
\end{equation}
with the mass operator
$$
\Sigma_{d;i\mu}(\varepsilon) = \sum_{\kappa}
V^*_{\kappa\mu}({\bf R}_i) g_{\kappa}(\varepsilon)
V_{\kappa\mu}({\bf R}_{i}) G_{d;i,i\mu}.
$$
which results from the exact solution of corresponding single
impurity problem for a given spin.\cite{Fleurov1,Fleurov2} Here the
index $\kappa = \{v,\nu\}$ unites both band and vacancy indices. The
bare propagators are defined as
$$
g_{v} = \frac{1}{\varepsilon -\varepsilon_v},\ \ g_{\nu} =
\frac{1}{\varepsilon - \varepsilon_\nu},\ \ g_{d;i\alpha} =
\frac{1}{\varepsilon - \varepsilon_{i\alpha}}
$$

Using Eqs. (\ref{Dyson1}) and (\ref{energy}) we can find the change
of the total energy due to the interaction of the vacancy and
impurity states. The second order correction in the vacancy -
impurity hybridization $V_{\mu,\nu}({\bf R}_i)$  in Eq.
(\ref{Dyson1}) is a single impurity effect and has nothing to do
with the exchange interaction. The latter appears only in the last
term of Eq. (\ref{Dyson1}), which is of the fourth order and
describes a shift of the vacancy levels due to the electron exchange
between two impurities, {\em which is possible only provided the
spins of two impurities are parallel}.

The upward shift of the empty vacancy levels corresponds to lowering
of the total energy of the system (see discussion after Eq.
(\ref{energy})). Hence, we come to the equation for the energy gain
due to an indirect exchange (superexchange) between the two magnetic
impurities,
\begin{equation}\label{exchange}
E_{ij}^{magn} = 2 \sum_{\nu\neq \nu'\ \mu\mu'} \frac{V^*_{\nu\mu}
({\bf R}_i) V_{\nu'\mu}({\bf R}_i) V^*_{\nu'\mu'}({\bf R}_j)
V_{v,\mu'} ({\bf R}_j) }{[\varepsilon_v - \varepsilon_{d;i\mu} -
P_{d;i\mu}(\varepsilon_\nu)] (\varepsilon_\nu - \varepsilon_{\nu'})
[\varepsilon_\nu' - \varepsilon_{d;j\mu'} - P_{d;j\mu'}
(\varepsilon_\nu')]}
\end{equation}
where $ P_{d,i\alpha}(\varepsilon) = \mbox{Re}
\Sigma_{d,i\alpha}(\varepsilon)$. Eq. (\ref{exchange}) has a
structure similar to that of eq. (10) of our paper \cite{Krst}. The
role of hole pockets is now played by the empty vacancy states, and
the fact that there is no free carriers and, hence no Fermi surface,
so important for the RKKY model, play now no role whatsoever.
Besides, contrary to the polaronic model\cite{Coey}, the
ferromagnetism with high enough $T_C$ arises in our model without
additional enhancement due to spin polarization of the vacancy
related band (see more detailed discussion  below).

The fourth order exchange energy can be interpreted by means of the
diagram presented in Fig. \ref{zen}. Now the dashed lines in this
diagram correspond to the two vacancy states virtually occupied in
the superexchange act. Two multipliers in the denominator in
(\ref{exchange}) are the energies $\widetilde\Delta_{do,\ i}$
required for the two "reactions" ${\rm Co}(d^7) \to {\rm Co}(d^6) +
V_O(p)$ in the sites $i,j$. These energies, include also the ligand
field shifts $P_{d;i\mu}$ of the impurity $d$
levels.\cite{Haldane,Fleurov1,Hald} The vacancy band is empty so
that the summation in Eq. (\ref{Dyson1}) is carried out over all the
states of the band. As a result the third multiplier,
$(\varepsilon_v - \varepsilon_{v'})$, in the denominator is of the
order of the vacancy band width $D_o$. Thus the calculation of
exchange energy by the Green function method by means of Eqs.
(\ref{energy}), (\ref{exchange}) confirms the qualitative estimate
(\ref{estim}).

Now we are in a position to derive the effective spin Hamiltonian
for interacting magnetic defects. Since we integrated out all the
charge degrees of freedom when calculating the energy
$E^{magn}_{ij}$, this Hamiltonian has the simple form
\begin{equation}\label{Heis}
H_{ex}^{eff}=\frac{1}{2}\sum_{\langle ij \rangle} J_{\langle ij
\rangle} {\bf S}_i{\bf S}_j~,
\end{equation}
where ${\bf S}_i$ is the spin of the ion ${\rm Co}(d^7)$, and the
exchange constant $J_{\langle ij \rangle}$ is determined as the
energy difference {\em per bond} between the parallel and
antiparallel orientations of spins in the pair $\langle ij \rangle$.
One may identify the coupling constants as $2 J_{\langle ij \rangle}
S(S+1) = E^{magn}_{ij}$. Since these coupling constants are
negative, they lead to a ferromagnetic ordering at temperatures
below the Curie temperature $T_C$. Well beyond the percolation
threshold (which may be identified with the minimal vacancy
concentration sufficient for formation of the vacancy band) $T_C$
may be estimated as
\begin{equation}\label{Curie}
T_{C}=  \frac{1}{k_B} \frac{z S(S+1)}{3}|J| = \frac{z}{6k_B}
|E^{magn}|.
\end{equation}
Here $|E^{magn}|$, and respectively $|J|$, is the typical value of
the superexchange interaction in the pair of impurities separated by
a distance not exceeding the double localization radius $R_O$ of the
vacancy state. $z$ is the number of Co impurities within the
corresponding volume. As was mentioned above the fact that 1\% of
vacancies is sufficient for formation of a band, meaning that the
typical distance between them is smaller than $2R_O$. Since nearly
each Co atom forms a complex with an oxygen vacancy, a 2\%
concentration of Co atoms substitutions means a 1\% vacancy
concentration when the vacancy states form a band. This fact
guaranties that each Co atom interacts not only with its own vacancy
but also with other neighboring vacancies, which form pairs with
other Co atoms.

Magnetic ordering with a high $T_C$ in ${\rm (Ti,Co)O_2}$ is
observed at higher Co concentrations up to 6\% or 3\% of the
concentration of vacancies captured by the Co impurities. A 1\%
concentration of vacancies creates a band in the forbidden energy
gap so that we may expect that the coordination number for such a
concentration may be at least 4. At a three times higher
concentration the coordination number will certainly become higher
and a rough estimate  $z\approx 10$ seems to be quite reasonable. We
do not dispose precise experimental data concerning positions of the
${\rm V_O}$ level in anatase TiO$_2$, so we refer the data available
for the rutile modification where the donor levels related to an
isolated $\rm V_O$ are found at a few tenth of 1 eV below the bottom
of conduction band. \cite{Halley} Using for the forbidden band width
the experimental value 3.2 eV and basing on the calculations of
Co-related d-level positions in the forbidden energy gap presented
in Ref. \onlinecite{Park}, one may estimate the $pd$ charge transfer
gap $\widetilde \Delta_{pd}\sim 1$eV. The width of the vacancy band
$D_o$ as well as the magnitude of hybridization parameter strongly
depend on the specific characteristics of the sample (vacancy
concentration, annealing regime etc). Besides, the vacancy band may
or may be not partially filled due to uncontrollable donor
impurities. Taking for an estimate the average value of
$V_{\nu\mu}\sim 0.1$ eV, assuming the same estimate for $D_o$, and
substituting the value of $S=3/2$ into Eq. (\ref{Curie}) one gets
$T_C \sim \gamma z\cdot 10^{-4}$eV $\approx \gamma 100$K for $z=10$.
The coefficient $\gamma$ includes all the uncertainties (degeneracy
factor of vacancy and impurity levels, difference between the
intracell and intercell hybridization parameters, uncertainties in
the effective coordination number $z$ etc. We expect the factor
$\gamma$ to be ideally close to one. However, it may deviate, and
even strongly, from this value depending on the preparation history
of each particular sample. Unfortunately, it is rather difficult to
find optimal conditions for high $T_C$ in such a multifactor
situation.

\section{Concluding remarks}

Among many dilute magnetic dielectrics only one example has been
chosen, namely oxygen deficient ${\rm (Ti,Co)O_2}$, and it has been
shown that the imperfection of this crystal is crucially important
for the formation of a long-range magnetic order. This idea is
supported by the recent experimental observation: ferromagnetism is
suppressed in Co doped ${\rm TiO_2}$ film with high structural
quality.\cite{Cr} Similar correlations between the quality of films
and the magnetism were discovered in Ref. \onlinecite{Cr} for ${\rm
(Ti,Cr)O_2}$. It follows directly from our theory that without
additional defects creating shallow levels under the bottom of
conduction band, the radius of magnetic correlation is too short to
overcome the percolation threshold, so that improvement of the film
quality is detrimental for magnetic ordering in DMD.

It is also noticed in Ref. \onlinecite{Cr} that the charge state of
chromium impurities in ${\rm TiO_2}$ is $\rm Cr^{+3}$. This fact is
easily understood within the framework of our hypothesis about the
origin of magnetic order in DMD. Indeed, the difference between Co
and Cr from the point of view of their electronic levels in host
${\rm TiO_2}$ [in fact, the addition energies defined in Eq.
(\ref{addit}) for the two impurities] is the position of these
levels relative to the shallow defect levels. In correlation with
the experimental data on ESR signal and theoretical calculations of
the addition energies,\cite{Mizu} the level of $\rm Cr^{2+}$ state
in rutile ${\rm TiO_2}$ is higher than the levels of shallow
impurities and apparently higher than the bottom of conduction band.
We assume that the situation is similar in anatase ${\rm TiO_2}$, so
that the same charge transfer mechanism, which binds the ions $\rm
Co^{2+}$ with oxygen vacancies, should stabilize chromium impurities
in a state $\rm Cr^{3+}$ (see Fig. 1). It is worth also mentioning
that generally we may expect that the chemical trends in the Curie
temperature for various transition metal impurities should correlate
with the deep level energies (e.g., \cite{klg06}).

Another experimental fact, which demands theoretical explanation
is a strong sample dependent scatter of saturation magnetic moment
$M_s$, which also depends on the fabrication
conditions.\cite{Coey,Cr} This fact may be easily explained if one
recognizes that only part of oxygen vacancies is bound to magnetic
impurities. "Free" vacancies, $V_O$, retain their electrons and
eventually donate them to the vacancy band. The partially occupied
band is spin polarized, and this polarization may enhance or
partially compensate the magnetization of transition metal ions
depending on the net sign of effective exchange in this
band.\cite{Anisim,Coey,Weng}

To conclude, we have proposed in this paper a mechanism of
ferromagnetic exchange in dielectric non-stoichiometric ${\rm
TiO_2}$ doped with transition metal impurities, which involves the
interaction between magnetic ions and oxygen vacancies and proposed
a qualitative explanation of the basic properties of these
materials. A quantitative calculations of electronic structure of
magnetically doped ${\rm TiO_2}$ based on the model described in
this paper will be presented in forthcoming publications.

\begin{acknowledgments}
The authors are indebted to D. Khomskii, K Krishnan, A. Pakhomov,
and T. Ziman for valuable comments.
\end{acknowledgments}


\begin{thebibliography}{99}
\bibitem{foot1} These Bloch tails are responsible for Co-$3d$
related peaks in the density of states near the top of the valence
band on ${\rm Ti_{1-x}Co_xO_2}$ in band calculations of Ref.
\onlinecite{Park}.
\end{thebibliography}
\end{document}